\documentclass[a4paper,11pt]{article}
\usepackage{graphicx}
\usepackage{graphics}
\title{\bf Target mass corrections and twist-3 in the nucleon 
spin structure functions}

\author{Y. B. Dong \\
Institute of High Energy Physics, 
Chinese Academy of Sciences,\\ 
Beijing 100049, P. R. China}
\begin{document}
%\date{}
\maketitle
\begin{abstract}

The Nachtmann moment is employed to study the contribution of twist-3 
operator to the nucleon spin structure functions. Target mass corrections to 
the Cornwall-Norton moments of the spin structure functions $g_{1,2}$ are 
discussed. It is found that the corrections play a sizeable role to the 
contribution of the twist-3 $\tilde d_2$ extracted from the Cornwall-Norton 
moments. 
 
\end{abstract}
\par
PACS: 13.60.Hb, 12.38.Aw, 12.38Cy; 12.38.-t; 13.60.Fz, 12.40.Nn
\par
Keywords: Target mass corrections; Nachtmann moment; Twist-3; Higher-twist.\\

\section{Introduction}

\par\noindent\par\vspace{0.2cm}

We know that the study of Bloom-Gilman quark-hadron duality is essential 
to understand the physics behind the connection between perturbative QCD 
(pQCD) and non-perturbative QCD [1]. In 2000, the new evidence of  
valence-like quark-hadron duality in the nucleon unpolarized structure 
functions $F_2^{p,d}$ was reported by Jefferson Lab. [2]. The well-known 
Bloom-Gilman quark-hadron duality [3] tells that prominent resonances do 
not disappear relatively to the background even at a large $Q^2$. It also 
means that the average of the oscillate resonance peaks in the resonance 
region is the same as the scaling structure function at a large $Q^2$ value. 
The origin of the Bloom-Gilman quark-hadron duality has been discussed by 
Rujula, Georgi and Politzer [4] with a QCD explanation. According to    
operator production expansion (OPE), it is argued that higher-twist 
effects turn to be small in the integral of the structure functions, 
and therefore, the leading-twist plays a dominate role to the moments of the 
nucleon structure functions [4]. So far, the nucleon structure functions and 
the higher-twist effects have already been carefully and systemically studied 
[5]. Some detailed calculations for the higher-twist effects were carried 
out based on various theoretical approaches, like bag model [6], QCD sum 
rule [7-8], constituent quark model [9], Lattice QCD [10], and chiral 
soliton model [11]. Moreover, there are also several empirical analyses 
of the spin structure functions of $g_1$ and $g_2$ at low $Q^2$. The 
higher-twist effects, like the ones of the twist-3 and twist-4 terms, 
have been extracted from the data [12-16]. Those analyses can be 
more and more accurate because more and more precisely new measurements 
of the nucleon spin structure functions of $g_1$ [17-19], and particularly of 
$g_2$ [15,16,20], are available.\\

Usually, the contribution of the twist-3  $\tilde d_2$ is extracted 
from the measured $g_{1,2}(x,Q^2)$ by calculating  the moment of   
\begin{eqnarray}
I(Q^2)=\int_0^1dx x^2\Big (2g_1(x,Q^2)+3g_2(x,Q^2)\Big )\rightarrow 
\tilde d_2(Q^2).
\end{eqnarray}
We know that the first moment of $g_1$ can be generally expanded 
in inverse powers of $Q^2$ in OPE [5]. It is 
\begin{eqnarray}
g_1^{(1)}=\int_0^1dx g_1(x,Q^2)=\sum_{\tau=2,even}^{\infty}
\frac{\mu_{\tau}(Q^2)}{Q^{\tau-2}}
\end{eqnarray}
with the coefficients $\mu_{\tau}$ related to the nucleon matrix elements 
of the operators of twist $\leq \tau$. In eq. (2), the leading-twist (twist-2) 
component $\mu_2$ is determined by the matrix elements of the axial vector 
operator $\bar{\psi}\gamma_{\mu}\gamma_5\psi$, summed over various quark 
flavors. The coefficient of $1/Q^2$ term contains the contributions from 
the twist-2 $\tilde a_2$, twist-3 $\tilde d_2$, and twist-4 
$\tilde f_2$, respectively. Thus [5], 
\begin{eqnarray}
\mu_4=\frac19M^2(\tilde a_2+4\tilde d_2+4\tilde f_2),
\end{eqnarray}
where $M$ is the nucleon mass. 
In eq. (3) $\tilde a_2$ arises from the target mass corrections, and it is of 
purely kinematical origin. It relates to the third  moment of the twist-2 
part of $g_1(x,Q^2;M=0)$ ($2\tilde a_2=\int_0^1x^2dx g_1(x,Q^2; M=0)$).
The other higher-twist terms, like $\tilde d_2$ and $\tilde f_2$, result 
from the reduced matrix elements, which are of the dynamical origin since 
they show the correlations among the partons [7,21]. If the contribution of 
the twist-3 term is well defined by eq. (1), then the contribution of the 
twist-4 term, like $\tilde f_2$, can be extracted 
according to eq. (3). \\

It should be mentioned that the method of eq. (1) to extract the twist-3 
contribution $\tilde d_2$ ignores the target mass corrections to $g_{1,2}$, 
since  the relations 
\begin{eqnarray*}
\int_0^1x^2g_1(x,Q^2)=\frac{1}{2}\tilde a_2,~~~~~
\int_0^1x^2g_2(x,Q^2)=\frac{1}{3}\Big (\tilde d_2-\tilde a_2\Big )
\end{eqnarray*} 
are used. In general, the 
nucleon target mass corrections should be considered completely in the 
studies of the nucleon structure functions [22], of the Bloom-Gilman 
quark-hadron duality [23-25], and of the Bjorken sum rule [26] at a moderate 
low $Q^2 (\sim 1-5~GeV^2$). We know that the target mass corrections 
to the nucleon structure functions are of the pure kinematical origin. They 
are different from the other higher-twist effects from dynamical multi-gluon 
exchanges or parton correlations. Before one can extract the higher-twist 
effects, it is important to remove the target mass corrections from 
the data [24]. There were several works about the target mass corrections to 
$F_{1,2}(x,Q^2)$ and $g_{1,2}(x,Q^2)$ in the literature [27-29]. Recently, 
the expressions of all the electromagnetic and electroweak nucleon spin 
structure functions with the target mass corrections have been explicitly 
given in Refs. [30-31]. \\

In this work, in order to precisely extract the contribution of the twist-3 
operators, the target mass corrections to eq. (1) will be discussed. In 
section 2, we explicitly give the target mass corrections to the integral 
$I(Q^2)$. Moreover, the advantage of the Nachtmann moments is stressed. In 
section 3, the numerical estimate of the target mass corrections to 
$I(Q^2)$ is given comparing to the result of the Nachtmann moment. The 
last section is devoted for conclusions. \\
 
\section{Twist-3 matrix elements and the target mass corrections}
\par\noindent\par\vspace{0.2cm}

Here, we use the notations of Piccione and Ridolfi [30] for the spin structure 
functions and for their moments. We know that the well-known Cornwall-Norton 
(CN) moments are
\begin{eqnarray}
g_{1,2}^{(n)}(Q^2) =\int_0^1 dx x^{n-1}g_{1,2}(x,Q^2).
\end{eqnarray}
In Refs. [30-31], the target mass corrections to the nucleon spin structure 
functions $g_1$ and $g_2$ are explicitly given in terms of the CN moments of 
the matrix elements of the twist-2 (leading-twist) operator and 
twist-3 one.  Up to twist-3, the results are 
\begin{eqnarray}
g_{1}^{(n)}(Q^2)&=&a_n+y^2\frac{n(n+1)}{(n+2)^2}\Big (na_{n+2}+4d_{n+2}\Big )
\nonumber \\
&&+y^4\frac{n(n+1)(n+2)}{2(n+4)^2}\Big (na_{n+4}+8d_{n+4}\Big )\nonumber \\
&&+y^6\frac{n(n+1)(n+2)(n+3)}{6(n+6)^2}\Big (na_{n+6}+12d_{n+6}\Big )
+{\cal O}\Big (y^8\Big )
\end{eqnarray}
with $y^2=M^2/Q^2$, and   
\begin{eqnarray}
g_{2}^{(n)}(Q^2)&=&\frac{n-1}{n}(d_n-a_n)+y^2\frac{n(n-1)}{(n+2)^2}
\Big (nd_{n+2}-(n+1)a_{n+2}\Big )\nonumber \\
&&+y^4\frac{n(n-1)(n+1)}{2(n+4)^2}\Big (nd_{n+4}-(n+2)a_{n+4}\Big )\nonumber \\
&&+y^6\frac{n(n-1)(n+1)(n+2)}{6(n+6)^2}\Big (nd_{n+6}-(n+3)a_{n+6}\Big )
+{\cal O}\Big (y^8\Big ).
\end{eqnarray}
In eqs. (5) and (6), $a_n$ and $d_n$ are the reduced hadron matrix elements 
of the irreducible Lorentz operators [29-30]:  
$R_1^{\sigma\mu_1\cdot\cdot\cdot\mu_{n-1}}$ and 
$R_2^{\lambda\sigma\mu_1\cdot\cdot\cdot\mu_{n-2}}$. The matrix elements
of the operators: $R_1^{\sigma\mu_1\cdot\cdot\cdot\mu_{n-1}}$ (twist-2) 
and $R_2^{\lambda\sigma\mu_1\cdot\cdot\cdot\mu_{n-2}}$ (twist-3)
can be written as  [29-30]
\begin{eqnarray}
<p,s \mid R_1^{\sigma\mu_1\cdot\cdot\cdot\mu_{n-1}}\mid p,s>=
-2Ma_nM_1^{\sigma\mu_1\cdot\cdot\cdot\mu_{n-1}},
\end{eqnarray}
and 
\begin{eqnarray}
<p,s \mid R_2^{\lambda\sigma\mu_1\cdot\cdot\cdot\mu_{n-2}}\mid p,s>=
 Md_nM_2^{\lambda\sigma\mu_1\cdot\cdot\cdot\mu_{n-2}},
\end{eqnarray}
where $M_1^{\sigma\mu_1\cdot\cdot\cdot\mu_{n-1}}$ is the general rank-n 
symmetric tensor which can be formed with one spin four-vector $s$ and 
$n-1$ momentum four-vectors $p$, and 
$M_2^{\lambda\sigma\mu_1\cdot\cdot\cdot\mu_{n-2}}$ is antisymmetric in  
$(\lambda, \sigma)$, symmetric in all other indices. The two tensors must be 
traceless. A typical example for a twist-3 operator is
\begin{eqnarray}
d_2:~~~~~\bar{\psi}\gamma_{\{\alpha}\tilde F_{\beta\}\gamma}\psi ~~~~~twist-3
 \end{eqnarray}
where $\{\cdot\cdot\cdot\}$ denotes symmetrizing the indices and  
subtracting the trace, and $\tilde F_{\alpha\beta}=
\frac12\epsilon_{\alpha\beta\gamma\delta}F_{\gamma\delta}$ is the dual 
gluon field strength [6-7]. We notice that the contribution of the 
leading-twist terms has a relation to the quark distributions functions. 
However, the contributions of the higher-twist  operators (like twist-3)
have no partonic interpretation [31]. In eq. (5) if we fix $n=1$, then, 
\begin{eqnarray}
g_1^{(1)}=a_1+y^2\frac{2}{9}(a_3+4d_3)+\cdot\cdot\cdot.
\end{eqnarray}
Comparing to eqs. (2-3), we see that $\tilde a_2=2a_3$ and $\tilde d_2=2d_3$. 
In eqs. (5) and (6) only the contributions of the leading-twist and 
twist-3 operators are considered.\\

From eqs. (1) and (4-6), we see that 
\begin{eqnarray}
I(Q^2)=2g_1^{(3)}+3g_2^{(3)}&=&\int_0^1x^2dx(2g_1(x,Q^2)+3g_2(x,Q^2))
\nonumber \\
&=&2\Big ( d_3+6y^2d_5+12y^4d_7+20y^6d_9\Big )+{\cal O}\Big (y^8\Big ) 
\nonumber \\
&=& \tilde d_2+6y^2\tilde d_4+12y^4\tilde d_6+20y^6\tilde d_8
+{\cal O}\Big (y^8\Big )
\end{eqnarray}
Clearly, $I(Q^2)$ contains the target mass corrections. It has mixed 
contributions from other higher-spin and twist-3 terms except for 
$d_3$(or $\tilde d_2$). \\

To get the twist-3 contribution of an operator with a definite spin, 
we need to consider the Nachtmann moments. According to Ref. [29], the 
two Nachtmann moments of the spin structure functions are defined as 
\begin{eqnarray}
M^{(n)}_1(Q^2)&=&\int^1_0dx\frac{\xi^{n+1}}{x^2}
\bigg \{\Big [\frac{x}{\xi}-\frac{n^2}{(n+2)^2}y^2x\xi
\Big ]g_1(x,Q^2)\nonumber \\
&-&y^2x^2\frac{4n}{n+2}g_2(x,Q^2)\bigg \}, 
~~(n=1,3,5,...),
\end{eqnarray}
and 
\begin{eqnarray}
M^{(n)}_2(Q^2)&=&\int^1_0dx\frac{\xi^{n+1}}{x^2}
\bigg \{\frac{x}{\xi}g_1(x,Q^2)\nonumber  \\
&+&\Big [\frac{n}{n-1}\frac{x^2}{\xi^2}-
\frac{n}{n+1}y^2x^2\Big ]g_2(x,Q^2)\bigg 
\}, ~~(n=3,5,...),
\end{eqnarray}
where the Nachtmann variable is [28]
\begin{eqnarray}
\xi=\frac{2x}{1+\sqrt{1+4y^2x^2}}.
\end{eqnarray}
Clearly, the two Nachtmann moments are simultaneously constructed by using 
the two spin structure functions $g_{1,2}$. According to  
eqs. (5-6), one can expand the two Nachtmann moments up to the order of 
$M^6/Q^6$ as
\begin{eqnarray}
M_1^{(n)}&=&a_n+ {\cal O}\Big (y^8\Big )\nonumber \\ 
M_2^{(n)}&=&d_n+{\cal O}\Big (y^8\Big ).
\end{eqnarray} 
Eq. (15) explicitly tells that, different from eqs. (1) and (11),  one can 
get the contributions of the pure twist-2 with spin-n and the pure 
twist-3 with spin-(n-1) operators from the Nachtmann moments. The above 
conclusion is valid when the target mass corrections are included. If the 
target mass corrections vanish, then eq. (11) turns to be $\tilde d_2$. The 
advantage of the Nachtmann moments, then, means that they contain only 
dynamical higher-twist effects, 
which are the ones related to the correlations among the partons. As a result, 
one sees that the Nachtmann moments are constructed to protect the moments 
of the nucleon structure functions from the target mass corrections. It is 
clear that the Nachtmann moments relate to the dynamical matrix elements of 
the operators with the definite twist and definite spin [17,21,24].\\

\section{The numerical estimate of the target mass corrections}

\par\noindent\par\vspace{0.2cm}

One may estimate the target mass corrections to the contribution 
of the twist-3 $\tilde d_2=2d_3$. We note that the nucleon spin structure 
functions are
\begin{eqnarray}
g_1(x,Q^2)&=&\frac{F_2(x,Q^2)[A_1(x,Q^2)+\gamma A_2(x,Q^2)]}
{2x[1+R_{\sigma}(x,Q^2)]},\nonumber \\
g_2(x,Q^2)&=&\frac{F_2(x,Q^2)[-A_1(x,Q^2)+A_2(x,Q^2)/\gamma]}
{2x[1+R_{\sigma}(x,Q^2)]},
\end{eqnarray}
where $F_2(x,Q^2)$ is the unpolarized structure function, 
$\gamma^2=4y^2x^2$, and 
$R_{\sigma}(x,Q^2)=\sigma_{L}(x,Q^2)/\sigma_T(x,Q^2)$ is 
the ratio of the longitudinal to transverse virtual photon cross sections. In 
eq. (16), $A_{1,2}(x,Q^2)$ are the virtual photon asymmetry parameters.  
To analyze the target mass corrections to the contribution of the twist-3 
$\tilde d_2=2d_3$ numerically, we simply employ the empirical 
parameterization forms of $g_1^{p,d}$ [16,18], of $R_{\sigma}(x,Q^2)$ [32], 
of $F_2^{p,d}$ [33], and of the average data of $g^{p,d}_2(x,Q^2)$ [20]. 
Here, we stress that the well-known Wandzura and Wilczek (WW) relation [34] 
for $g_2$ 
\begin{eqnarray}
g_2(x,Q^2)=g_2^{WW}(x,Q^2)=-g_1(x,Q^2)+\int_x^1\frac{g_1(y,Q^2)}{y}dy
\end{eqnarray}
is valid [30-31] if only the leading-twist is considered. The target mass
corrections to the twist-2 contribution do not break the WW relation. 
If the higher-twist operators, like twist-3, are considered, the WW relation 
$g_2(x,Q^2)=g_2^{WW}(x,Q^2)$ is no longer valid. Thus, according to Ref. [20], 
one may write  
\begin{eqnarray}
g_2(x,Q^2)=g_2^{WW}(x,Q^2)+\bar{g}_2(x,Q^2).
\end{eqnarray}
Clearly, if only the twist-2 term is considered, $\bar{g}_2(x,Q^2)=0$. The 
non-vanishing value of $\bar{g}_2$ results from the higher-twist effects. 
In Ref. [20], the average data of $A_2(x_0,Q_0^2)$ and 
$xg_2^{p,d}(x_0, Q_0^2)$ are listed for the  measured values of $x_0$ and 
$Q_0^2$. The measured ranges of the experiment are $0.021\leq x_0\leq 0.790$ 
and $0.80~GeV^2\leq Q_0^2\leq 8.20~GeV^2$. It is argued that 
$\bar{g}_2(x,Q^2)$ is independent of $Q^2$ in the measured region since  
$\tilde d_n$ depends only logarithmically on $Q^2$ [5]. Thus, the 
$Q^2$-dependence of the structure function $g_2(x,Q^2)$ mainly results from 
$g_2^{WW}(x,Q^2)$ [20]. Another point made by E155 collaboration [20] is 
that the contributions of $\bar{g}_2$ to the integral of eq. (1) beyond 
the measured $x$ region, say $x<0.021$ and $x>0.790$, are negligible. \\

Based on the above two arguments of Ref. [20], we calculate the ratio 
$R(Q^2)$ of the Nachtmann moment to the value of eq. (1)
\begin{eqnarray}
R(Q^2)=\frac{2M_2^{(3)}(Q^2)}{I(Q^2)}
\end{eqnarray}
for the proton and neutron spin structure functions, respectively. For the 
neutron structure function, we simple use the relation
\begin{eqnarray}
g_1^n(x,Q^2)=\frac{2}{1-\omega_D}g_1^d(x,Q^2)-g_1^p(x,Q^2),
\end{eqnarray}
with $\omega_D\sim 0.05$ being the deuteron D-state probability. 
It should be mentioned that the target mass corrections to the deuteron 
(spin one) structure fucntions have been discussed in Ref. [35]. It is found 
that the target mass corrections to $F_{1,2}^d$ and $g_{1,2}^d$ are precisely 
the same as in the spin-1/2 targets.  The 
calculated results of $R(Q^2)$ are plotted in Fig. 1. Clearly, the 
divergence of the two ratios from unity displays the target mass corrections. 
From Fig. 1 we see a sizeable role played by the target mass corrections. 
When the momentum transfer $Q^2$ becomes small, the role increases 
obviously. The effect of the target mass corrections is about 30\% for the 
average of the proton and neutron targets when 
$Q^2\sim 1.6~ GeV^2$. It decreases to about 10\% when $Q^2\sim 8~GeV^2$. 
We find that the values of the Nachtmann moment $2M_2^{(3)}$, 
which contains no target mass corrections, are always smaller than those 
of $I(Q^2)$ which has the target mass corrections. Therefore, we conclude 
that the contribution of the twist-3 $\tilde d_2$ to the nucleon structure 
function extracted from eq. (1) is overestimated. For example, at an average 
$Q^2$ of $5~GeV^2$ of the E155 experiment [20], the estimated 
$\tilde d_2^p=0.0025\pm 0.0016\pm 0.0010$ from experiment data and eq. (1). 
If the target mass corrections are taken into account, the central value of 
$\tilde d_2^p$ is reduced to about $0.00215$. This new value becomes 
even close to zero. Moreover, Fig. 1 shows that the ratio of the neutron is 
similar to that of the proton, and  the target mass corrections to the 
neutron is only a little bit smaller than those to the proton. In addition, 
the uncertainties of $\tilde d_2$ in Ref. [20] come from the statistical and 
systematic errors on measured $xg_2$. At a low $Q^2$ value, say, 
$Q^2=1~GeV^2$, the measured $xg_2^p$ is $0.010\pm 0.007$. When $Q^2$ 
increases to about $8~GeV^2$, the measured $xg_2^p=-0.007\pm 0.002$. Clearly, 
the errors on $\tilde d_2$ caused by the uncertainties of $xg_2$ are much 
larger than the target mass corrections in our calculation. \\

{\hskip 1in}
\begin{figure}[t]
\centering
\includegraphics[width=4in,height=3in]{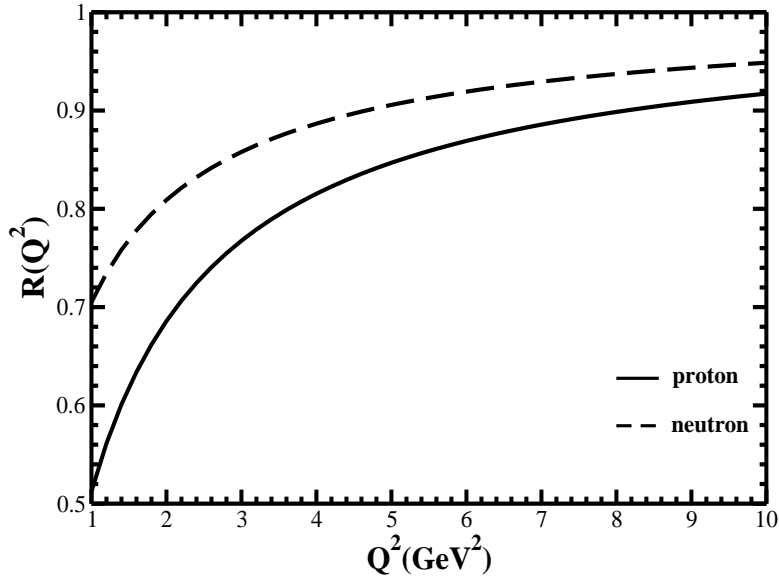}
\caption{\footnotesize Ratio $R(Q^2)$. The solid and dashed curves are 
the results of the proton and neutron, respectively. }
\end{figure} 

\section{Conclusions}
\par\noindent\par\vspace{0.2cm}

In summary, we have explicitly shown the target mass corrections to the 
contribution of twist-3 extracted from eq. (1). It is clear that 
the Nachtmann moments are constructed by the two spin structure functions and 
do not governed by  the target mass corrections. Therefore, in order to 
precisely extract the contribution of the twist-3 $\tilde d_2$ at different 
$Q^2$ values, one is required to employ the Nachtmann moment $2M_2^{(3)}$. 
The numerical estimate ratios $R(Q^2)$ for the proton and neutron 
in Fig. 1 indicate that the target mass corrections play a sizeable role 
to the integral of eq. (1). Thus, the equation gives the result which is 
different from the pure contribution of the twist-3 $\tilde d_2$. Moreover, 
the values of $\tilde d_2$ from eq. (1) are always overestimated 
because $I(Q^2)$ contains the mixed contributions of other twist-3 terms with 
higher-spin, except for $d_3$ (or $\tilde d_2$). Numerically, the effect  
of the target mass corrections in the extracted $\tilde d_2$ (or $d_3$) from 
eq. (1) is about 30\% for the average of the proton and neutron targets 
when $Q^2\sim 1.6~GeV^2$ or about 10\% when 
$Q^2\sim 8~GeV^2$. The corrections to the neutron are similar to those to the  
proton. As a result, we expect that one should take the Nachtmann 
moment to precisely extract the contribution of twist-3 operator $\tilde d_2$. 
With a very accurate value of $\tilde d_2$, one may further get the precise 
information of the other higher-twist operators, like twist-4 term 
$\tilde f_4$.  Since the errors of the present data of $xg_2^{p,d}$ 
are large, the uncertainties of the extracted values of 
$\tilde d_2^{p,d}$ from eq. (1) are larger than the target mass corrections. 
A much more precise measurement of $g_2$ and $A_2$ is urgently required. \\
     
\section*{Acknowledgments}
\par\noindent\par\noindent\par 

This work is supported by the National Sciences Foundations of China
under grant No. 10475088, by the CAS Knowledge Innovation Project 
No. KJCX3-SYW-N2, and by the Center of Theoretical Nuclear Physics, National 
Lab. of Heavy Ion Accelerator, Lanzhou. Communications with S. Rock are 
acknowledged.\\

\end{document}